\documentstyle[prl,preprint,aps,epsf]{revtex}
\begin{document}
 \draft
 \preprint{GTP-98-03}
  \title{Escape rate of the nanospin system in a magnetic field:
         the first-order phase transition within quantum regime}
 \author{Chang Soo Park$^1$, Sahng-Kyoon Yoo$^2$, D. K. Park$^3$ and Dal-Ho Yoon$^4$}
 \address{$^1$Department of Physics, Dankook University Cheonan 330-714, Korea
          \\
          $^2$Department of Physics, Seonam University, Namwon, Chunbuk 590-711,
          Korea \\
          $^3$Department of  Physics, Kyungnam University, Masan 631-701, Korea
          \\
          $^4$Department of Physics, Chongju University, Chongju 360-764, Korea}
 \date{today}
 \maketitle

 \begin{abstract}
  We have investigated the escape rate of the nanospin
  particle with a magnetic field applied along the easy axis.
  The model studied here is described by the Hamiltonian $\hat{\cal H}
  = K_1 \hat{S}_z^2 + K_2 \hat{S}_y^2 + g\mu_b H \hat{S}_x $, $(K_1 > K_2 > 0)$ from which
  the escape rate is calculated within the
  semiclassical approximation. We have obtained a diagram for the
  orders of the phase transitions depending on the anisotropy constant and
  the external field. For $ K_2 / K_1 > 0.85$ the
  present model reveals, for the first time, the existence of the
  first-order transition within the quantum regime.
 \end{abstract}

In recent years there has been much interest in the problem of the
magnetization reversal of a nanospin particle\cite{stamp}. It is well known
that there are two possible mechanisms of the magnetization
reversal : a classical thermal activation\cite{neel} and a quantum
tunneling\cite{ya,enz,chud88}. In the study of this problem the transition rate of the
system of magnetic particles is mainly concerned. At high
temperature the transition rate is governed by the classical thermal
activation, but when the temperature is
very low the quantum tunneling dominates. As the temperature is
lowered the phase change of the transition rate between the classical
thermal activation and the quantum tunneling occurs. This phase
transition can be either first-order\cite{larkin} or
second-order\cite{aff}. In recent several
works\cite{chud97,liang,lee} these are shown to be possible in the real system
such as single-domain ferromagnetic particle. These two types of
phase transitions have been suggested by Chudnovsky\cite{chud92}. However, the
coexistence of the first-order phase transition within quantum
regime and the second-order classical-to-quantum transition, which
was also proposed by Chudnovsky, was not found yet. In this letter we
find a real system which shows all kinds of phase transitions
mentioned above.

Consider a nanospin particle with an applied field ${\bf H}$ along
the easy axis. If the spin particle is a uniaxial
spin system with $XOY$ easy plane anisotropy and the easy $X$-axis
in the $XY$-plane the Hamiltonian for this system is given
by\cite{chud88}
\begin{equation}
\hat{\cal H} = K_1 \hat{S}_z^2 + K_2 \hat{S}_y^2 - g {\mu}_b H
\hat{S}_x
\end{equation}
where $K_1 , K_2$ are the anisotropy constants, ${\mu}_b$
is the Bohr magneton, and $g$ is the spin $g$-factor
which is taken to be 2.0 here. Since we
choose $XY$-plane as the easy plane the anisotropy constants
satisfies $K_1 > K_2 > 0$.

The anisotropy energy associated with this Hamiltonian has two
local energy minima ; the one on the $+X$-axis which is metastable
state and the other on the $-X$-axis. Between these two energy
minima there exists an energy barrier, and the spin escapes this
metastable state either by crossing over or by tunneling through
the barrier.

In the coherent spin state representation\cite{man} the effective Lagrangian
corresponding to the Hamiltonian, Eq.(1), for small deviation from the easy plane can
be written as
\begin{eqnarray}
{\cal L}(\phi, \dot{\phi}) &=& \dot{\phi} \hbar S_z - \langle \theta, \phi
\vert \hat{\cal H} \vert \theta, \phi \rangle \nonumber \\
&=& \hbar^2 \frac{m(\phi)}{2} {\dot{\phi}}^2 - V(\phi)
\end{eqnarray}
where
\begin{equation}
m(\phi) = \frac{1}{2K_1 (1 - \lambda \sin^2 \phi - \frac{\alpha
\lambda}{2} \cos \phi )}
\end{equation}
is an effective mass, and
\begin{equation}
V(\phi) = K_2 S^2 (\sin^2 \phi + \alpha \cos \phi + \alpha )
\end{equation}
is an effective potential for the spin system . Here we have
introduced dimensionless parameters $\lambda = K_2 / K_1 (<1)$, $\alpha =2 H/H_c$
($H_c = \frac{K_2 S}{\mu_b}$ being the coercive field), and
added a constant term $K_2 S^2 \alpha$ in the effective potential
for convenience. The effective mass and potential are shown in Fig.1. The
barrier height of $V(\phi)$ decreases as $\alpha$
increases and vanishes at $H=H_c$ for a given $\lambda$. Thus, in order for the
tunneling to exist $\alpha$ should have values between 0 and 2.
We note that the mass depends on $\phi$. Later we will see that the
$\phi$ dependence of the effective mass plays a crucial role for
the occurrence of the first-order phase transition in quantum
regime.

At temperature $T$ the escape rate of the spin particle can be
obtained by taking ensemble average of the tunneling probability.
Introducing Euclidean-time $\tau=i t$ this can be written as the
path integral form\cite{langer,callen,fey}
\begin{equation}
\Gamma (\tau) = \int d [\phi(\tau)] {\rm e}^{-\frac{1}{\hbar}
\oint d\tau [\hbar^2 \frac{m(\phi)}{2} {\dot{\phi}}^2 - V_E
(\phi)]} ,
\end{equation}
where $V_E (\phi) = -V(\phi)$ is the Euclidean effective potential
(see Fig.1), and $\dot{\phi} \equiv d \phi /d\tau$. In the
semiclassical approximation, neglecting the quantum fluctuation
term, the escape rate at an energy $E$ above the metastable
minimum is given by
\begin{equation}
\Gamma (\tau) \sim {\rm e}^{- \frac{1}{\hbar} S(\tau)} ,
\end{equation}
where $S(\tau)$ is the minimum effective Euclidean action which can
be obtained by taking the smallest value of $S_0$ and $S(T)$.
Here, $S_0$ is the thermodynamic action defined by
\begin{equation}
S_0 = \frac{\hbar E_0}{k_B T}
\end{equation}
with $E_0 = \frac{K_2 S^2}{4} (\alpha + 2)^2$, and $S(T)$ is
expressed as
\begin{equation}
S(T) = 2\hbar \int_{\phi_i (E)}^{\phi_f (E)} d \phi {\sqrt{
2m(\phi)[V(\phi)-E]}} + \frac{E\hbar}{k_B T} ,
\end{equation}
where ${\phi_i}(E)$ and ${\phi_f}(E)$ are the solutions of the equation
$V(\phi) = E$. For $E=2\alpha K_2 S^2$(the bottom of the
metastable state) the Euler-Lagrange equation gives the bounce
solution. When $2\alpha K_2 S^2 < E < \frac{K_2 S^2 }{4} (\alpha + 2)^2$
the trajectory $\phi(\tau)$ in $V_E (\phi)$ shows periodic motion with turning
points at $\phi_i (E)$ and $\phi_f (E)$. The solution
corresponding to this trajectory is called the periodic instanton
whose period is defined as
\begin{equation}
\tau_p (E) \equiv \frac{\hbar}{k_B T} = \hbar \int_{\phi_i
(E)}^{\phi_f (E)} d\phi \frac{\sqrt{2m(\phi)}}{\sqrt{V(\phi) - E}} .
\end{equation}

We now examine how the period $\tau_p$ changes as a function of
energy $E$. Since the effective mass depends on $\phi$ it
influences on the variation of $\tau_p$ with $E$. To see this we look
into the Eq.(3). For small values of $\lambda$, since
$m(\phi)$ varies not much it gives little effect on the behavior of
$\tau_p$. However, as $\lambda$ comes close to 1, the
magnitude of the effective mass at turning points is small
at first, then rises rapidly (see Fig.1). As the mass
becomes larger the speed of a particle in the potential $V_E (\phi)$
reduces, and it takes more time to complete the periodic motion in
$V_E (\phi)$. Thus, the period $\tau_p$ decreases with $E$ at the
start, but then changes to increase due to the rapid increase of
$m(\phi_f )$. We now note that the maximum point of the effective
mass $\phi_m$ does not
coincide with the minimum point of the Euclidean potential $\phi_0$,
which is illustrated in Fig.1. Therefore, as $E$ approaches to the
top of the potential barrier the motion of the particle in $V_E (\phi)$ is restricted
in a region where the effective mass becomes small. Thus, in this region,
$\tau_p$ decreases with $E$. This suggests that the whole behavior
of $\tau_p$ will be the form illustrated in Fig.2. As proposed in
Ref.\cite{chud92} this form produces the first-order phase transition inside
the quantum tunneling region.

In Fig.3 we have plotted the effective action  as a
function of $T$ for $\alpha = 1$ and $\lambda = 0.9$. At $T = T_c$ the
escape rate changes from the thermal activation to the
quantum tunneling regime, and the transition is second-order.
Below $T_c$, as $T$ is lowered, the minimum action
increases smoothly at first, changes abruptly at $T_{qc}$
(the cusp at $T = T_{qc}$ in the figure), and then becomes almost
constant. The quantum regime is, thus, divided into two parts :
the thermally assisted quantum tunneling (when $T_{qc} < T < T_c$)
and the pure quantum tunneling (when $T < T_{qc}$).
It can be seen from this picture that the phase transition at
$T = T_{qc}$ is first-order.

According to our numerical calculations the orders of the phase
transitions are relevant to both $\alpha$ and $\lambda$. In Fig.4
we have drawn a diagram for the orders of the phase transitions in
$(\lambda, \alpha)$ plane.
As remarked earlier the maximum value of $\alpha$ is 2
in the spin system. However, since we are interested in the
positive effective mass $\alpha$ is restricted by the inequality
\begin{eqnarray}
\lambda (1 + \frac{\alpha^2}{16} ) < 1.
\end{eqnarray}

From the diagram we observe many interesting results. First, the
classical-to-quantum phase transition shows both the first-order
(region I) and the second-order (region II) transitions. Note that
there is only the second-order transition for $\lambda < 0.5$. For
materials with $\lambda$ larger than 0.5 we can see that the order
changes from first to second as $\alpha$ increases, and the phase
boundary increases with $\lambda$ up to
0.85, after which it decreases.

In the case of the phase transition within the quantum regime
there is no phase transition
for $\lambda$ below 0.85. However, for $\lambda > 0.85$, there
exists phase transition which is first-order.
We also observe that for $0.85 < \lambda < 0.91$
the phase boundary starts from the value corresponding
to the maximum of the phase boundary between regions I and II and
increases with $\lambda$. When $\lambda$ becomes larger than 0.91,
however, the phase boundary decreases with $\lambda$ due to the positive
mass condition, Eq.(9). This phase boundary forms a new region III in
which both the first-order transition within the quantum regime and the second-order
classical-to-quantum transition coexist. Finally, in the region
IV, since the negative effective mass begins to appear the phase transition
cannot be defined.

Our speculation about
these results is as following. For materials with small $\lambda$
the height of the potential
barrier is small. In an ensemble of nanospin particles each will
then be relaxed to reverse its magnetization easily, which leads to the
smooth variation of the escape rate with temperature $T$. On the
other hand, if $\lambda$ is large the barrier height will become
large, and hence the spin particle will be reluctant to reverse its
magnetization. In this case it needs the energy such as latent heat
to make the spin particles
ready to reverse their magnetizations. Therefore, the escape rate
experiences the first-order phase transition.

We can also discuss the change of the order with $\alpha$ for a
given $\lambda$. For $\lambda$ less than 0.5, since the barrier
height is essentially small for all values of $\alpha$, the
nanospin particle can easily reverse its magnetization, which
corresponds to the second-order transition. For $0.5 < \lambda <
0.85$, the barrier height is large at small $\alpha$, but it
becomes smaller as $\alpha$ increases. It is thus possible for the order
to change from first to second with increasing $\alpha$. We now consider
the case $\lambda > 0.85$. When $\alpha$ is small the barrier
height is so large that only the high temperature first-order
classical-to-quantum transition is possible. However, when $\alpha$
is moderately large, the first-order transition occurs at lower
temperature, i.e., quantum region, with the classical-to-quantum
transition being changed into second-order. This leads to the
first-order quantum transition in quantum tunneling region. As $\alpha$
further increases, the external field makes the magnetization
reversal easy (small barrier height), the situation is same as the
case of small $\lambda$.

The Fig. 5 represents the crossover temperatures as a function of $\alpha$
for $\lambda = 0.9$. It tells that both $T_c$ and $T_{qc}$ decrease
as $\alpha$ gets large. This is rather obvious from the fact that
the depth of the metastable well becomes shallow with increasing
$\alpha$.

In conclusion, we have investigated the phase transition of the escape rate
from metastable states in nanospin system with a magnetic field applied along the
easy axis. We found the coexistence of the first-order phase
transition within the quantum tunneling region and second-order
classical-to-quantum transition for large  $\lambda$ and
$\alpha$, which had not been observed before.
Furthermore, the phase diagram for the orders of the phase transitions in
$(\lambda , \alpha)$ plane is obtained. This phase diagram can be
used as a guide for the experimental observations.

 \newpage
 \begin{figure}
 \caption{The effective potential $V(\phi)$(solid), Euclidean potential $V_E
 (\phi)$(dashed), and the effective mass $m(\phi)$(dotted). $\phi_0$ is the
 position at which the Euclidean potential has a minimum, while $\phi_m$
 the position at which the effective mass has a maximum. $\phi_i (E)$
 and $\phi_f (E)$ are the classical turning points at Euclidean
 energy $-E$. }
 \label{fig1}
 \end{figure}

\begin{figure}
\caption{ The period in Euclidean potential as a function of
energy at $\lambda = 0.9$ and $\alpha = 1.0$, which shows the first-order
phase transition within the quantum regime.}
\label{fig2}
\end{figure}

\begin{figure}
\caption{ The actions $S(T)$ and $S_0$ as a function of
temperature at $\lambda = 0.9$ and $\alpha = 1$.
$T_c$ corresponds to classical-to-quantum crossover
temperature and $T_{qc}$ to the transition temperature between the
different quantum regime. }
\label{fig3}
\end{figure}

\begin{figure}
\caption{ The phase diagram for the orders of phase transition in $(\lambda, \alpha)$
plane. Region I: the first-order classical-to-quantum transition. Region
II: the second-order classical-to-quantum transition. Region III:
the second-order classical-to-quantum transition and the first-order
transition within quantum regime coexist. Region IV: the negative
effective mass area. }
\label{fig4}
\end{figure}

\begin{figure}
\caption{ Crossover temperature as a function of $\alpha$. Both $T_c$
and $T_{qc}$ decrease with increasing $\alpha$. }
\label{fig5}
\end{figure}

\newpage
\epsfysize=15cm \epsfbox{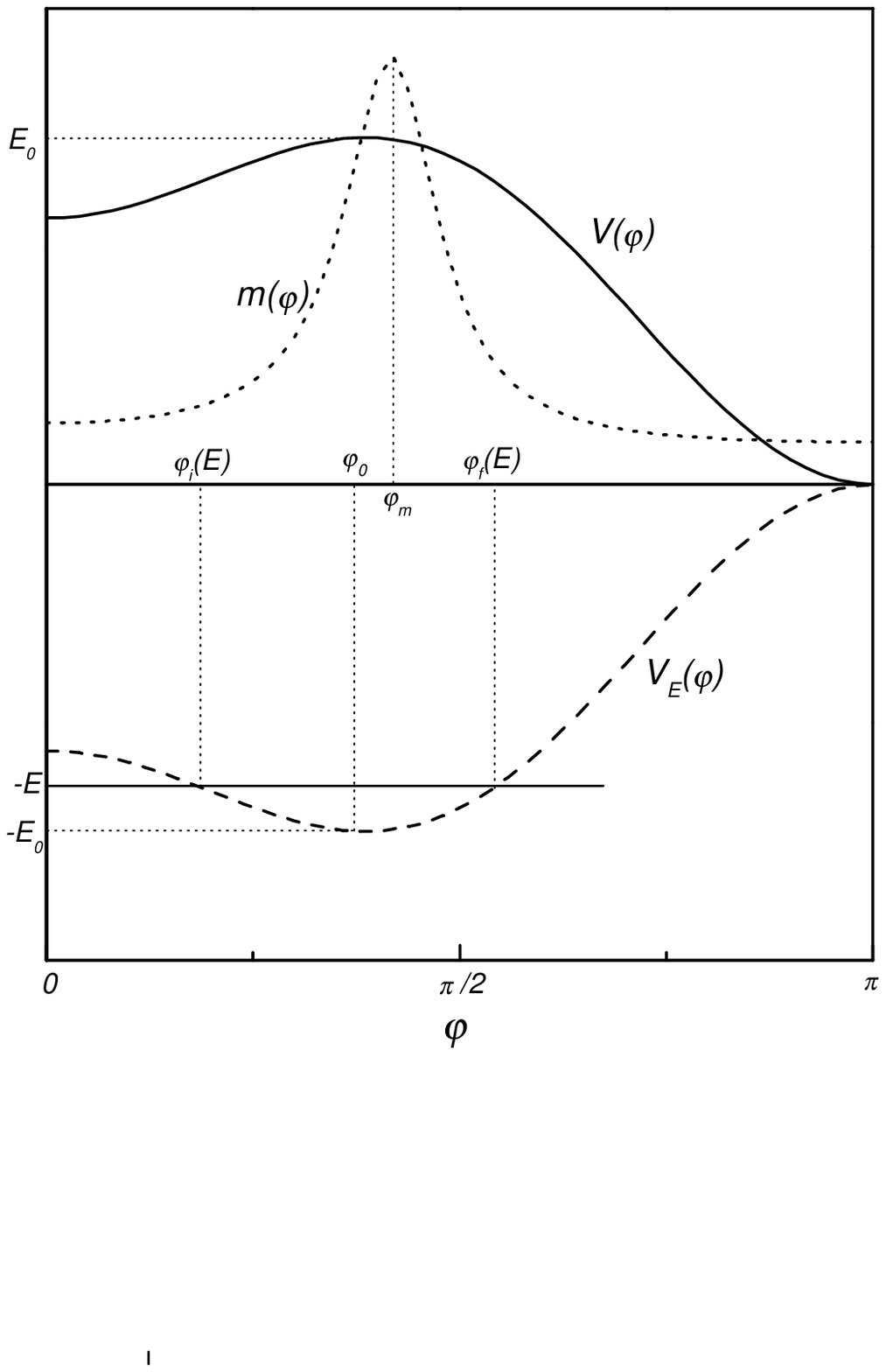}

\newpage
\epsfysize=15cm \epsfbox{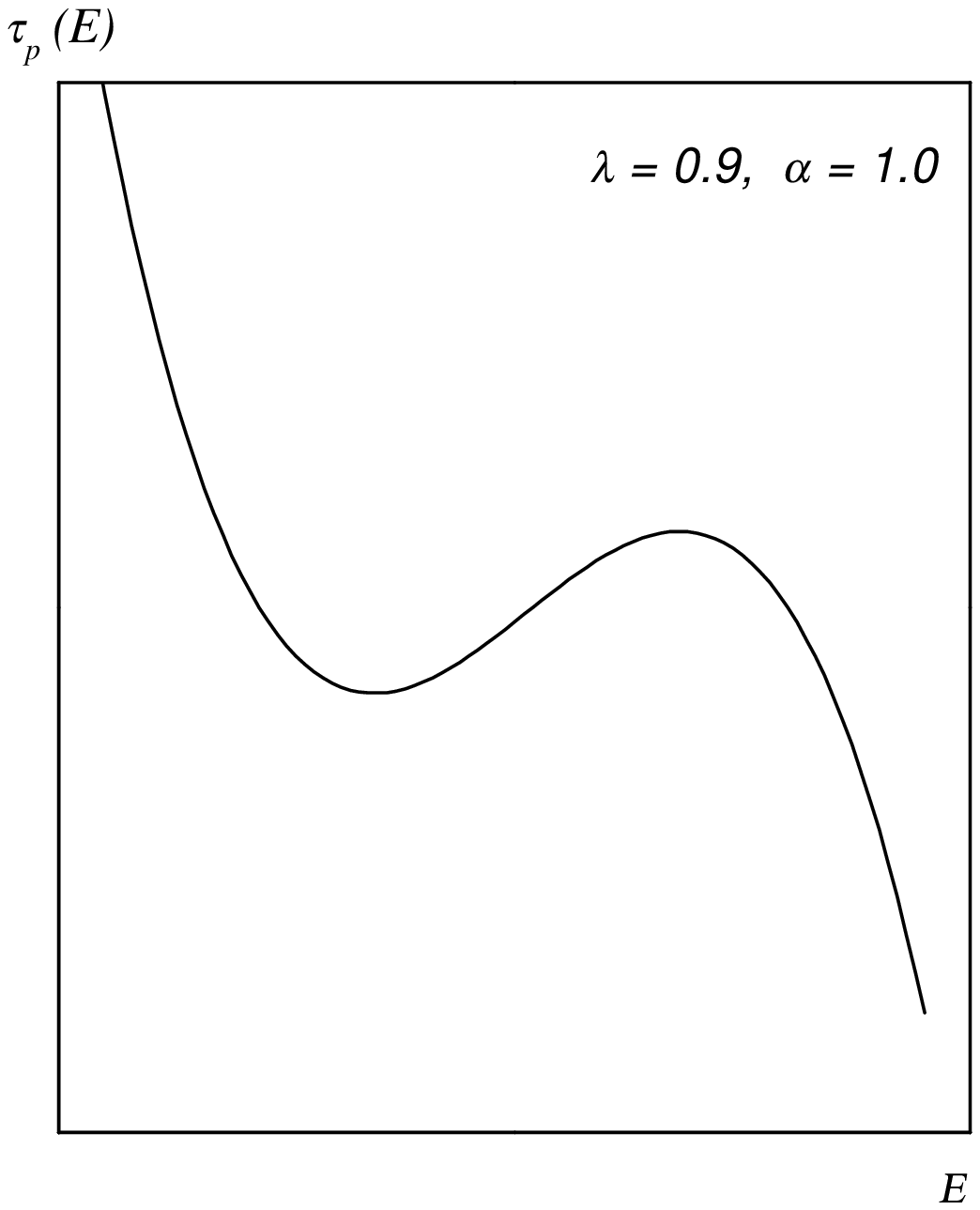}

\newpage
\epsfysize=15cm \epsfbox{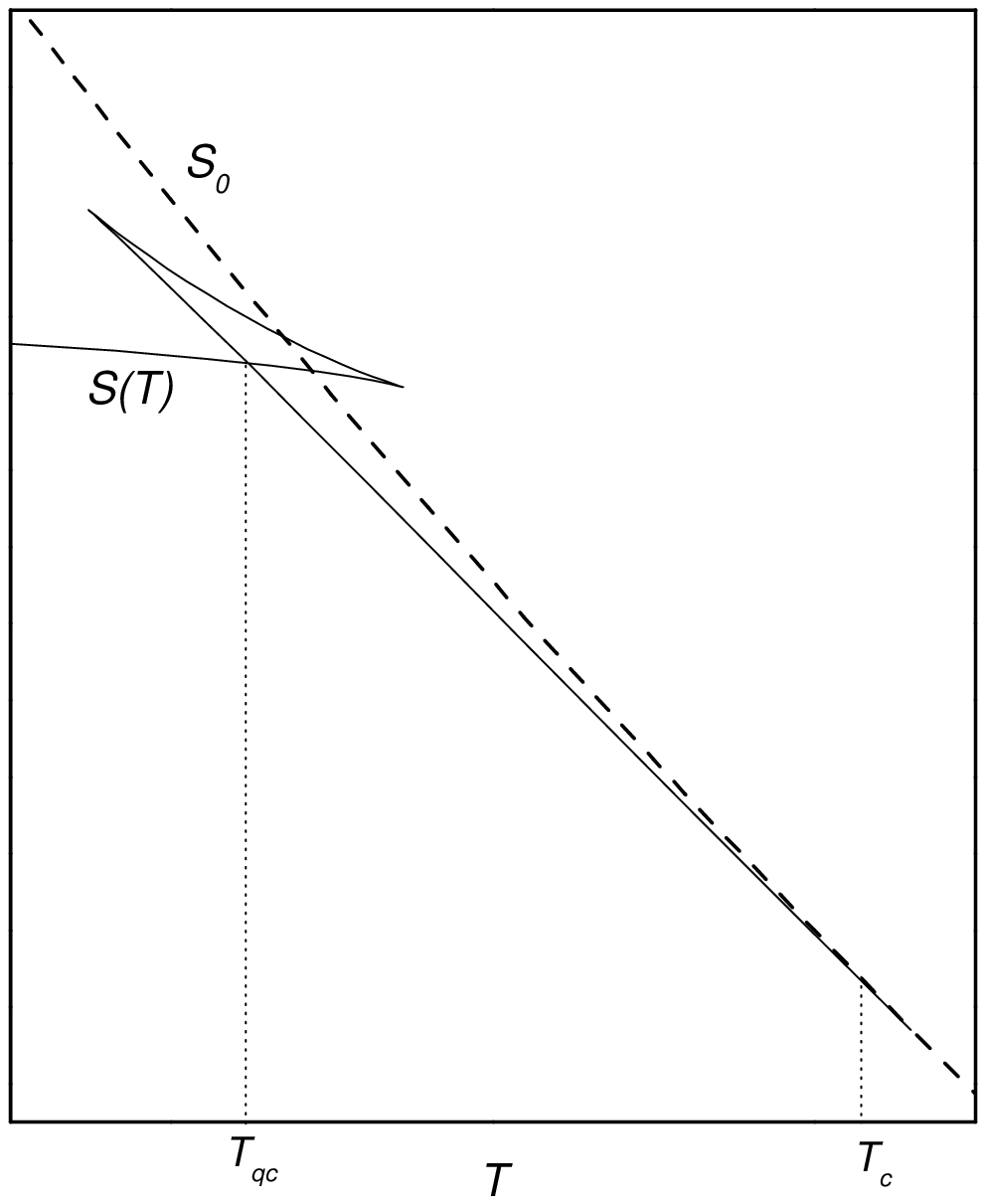}

\newpage
\epsfysize=15cm \epsfbox{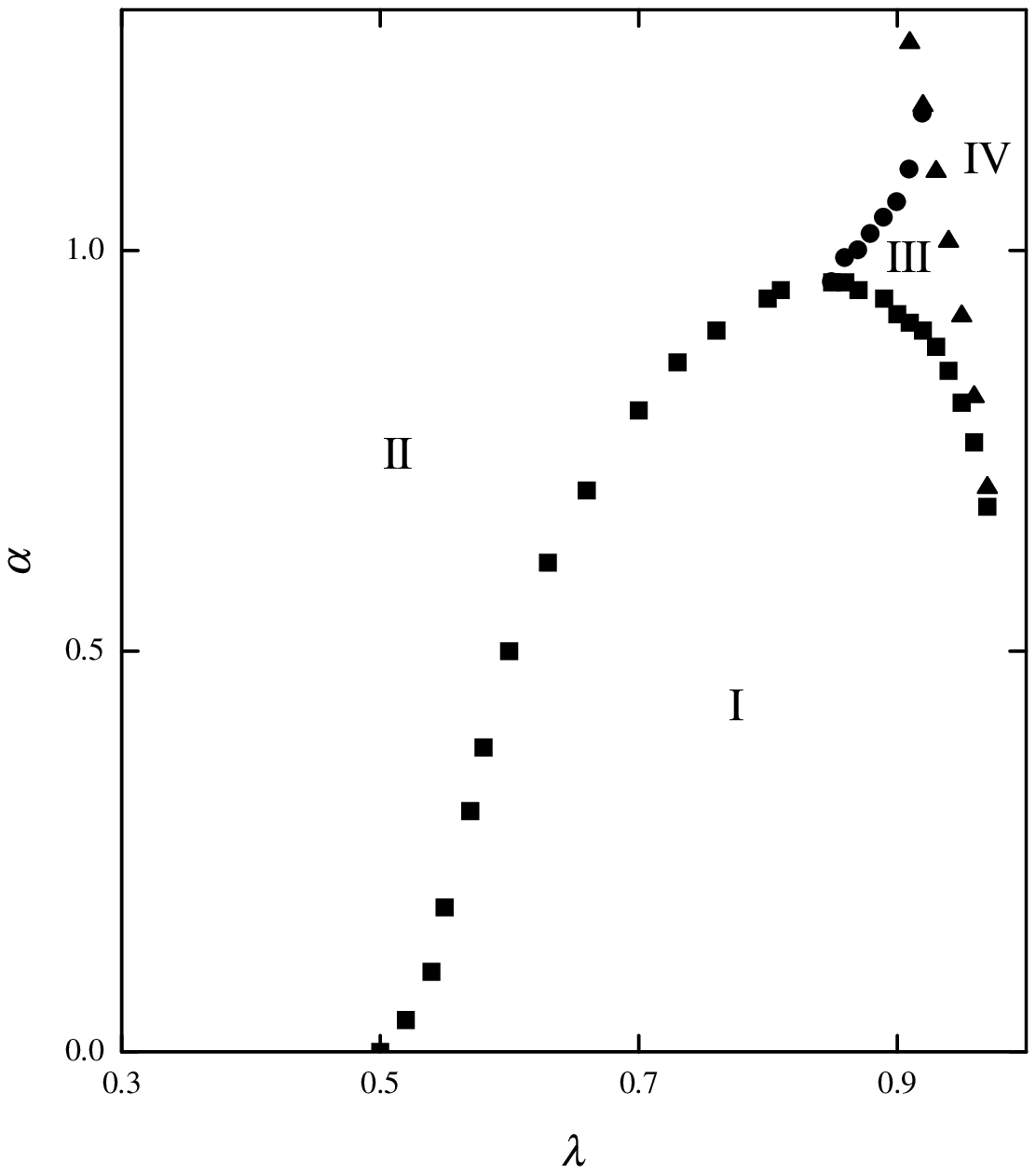}

\newpage
\epsfysize=15cm \epsfbox{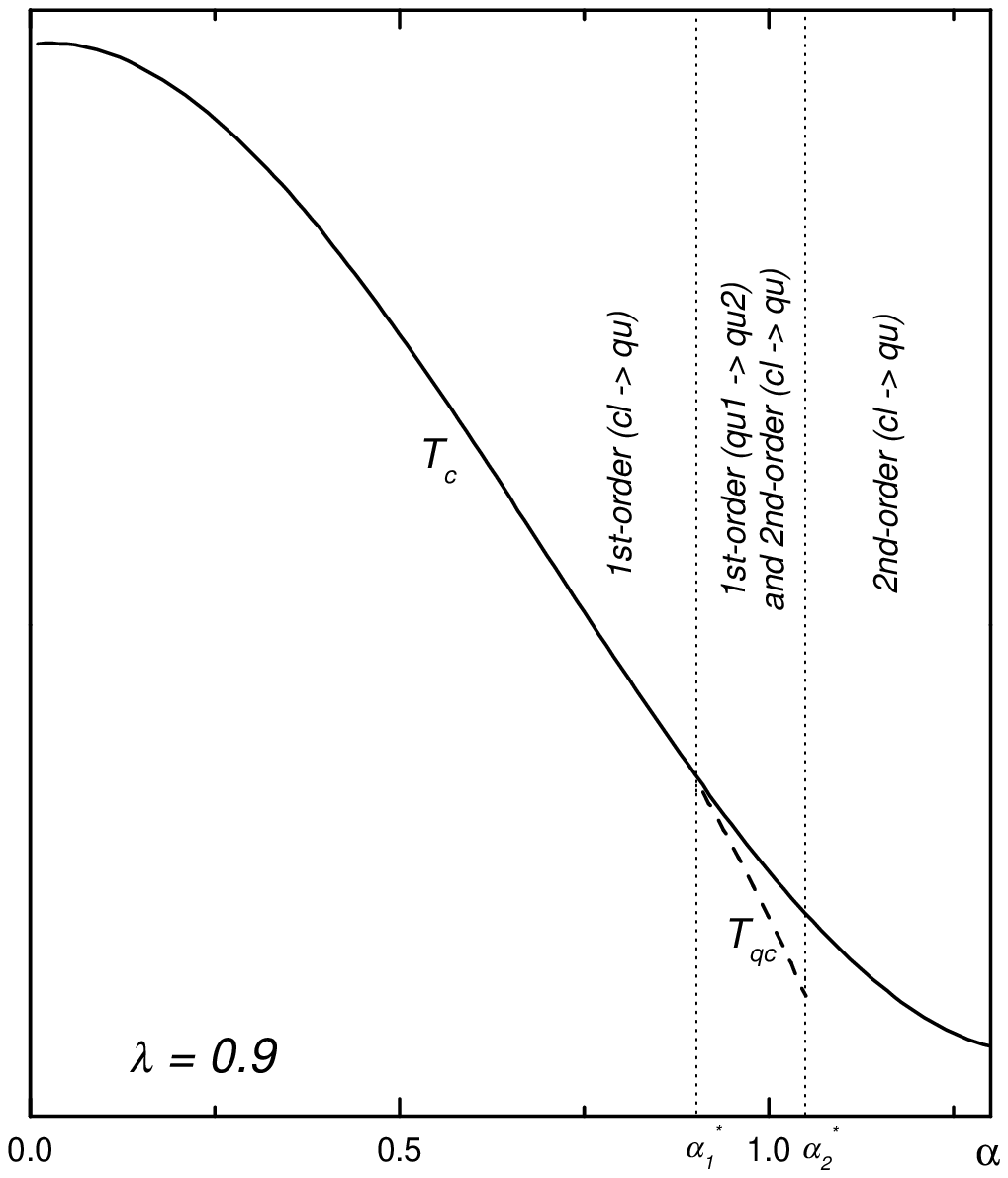}

\end{document}